\documentclass[aps,prl,twocolumn,final,letterpaper]{revtex4}

\usepackage{graphicx}   
\usepackage{import}                         
\usepackage{epstopdf}
\usepackage{amsmath} 
\usepackage{bm}
\usepackage{amssymb}
\usepackage{quotes}
\usepackage{transparent}
\usepackage{dcolumn}
\usepackage[usenames, dvipsnames]{color}
\usepackage{multirow}
\usepackage{cancel} 
\usepackage{mdframed}
\usepackage{color}
\usepackage{bm}
\usepackage{setspace}
\usepackage{dsfont}
\usepackage{braket}
\usepackage{slashed}
\usepackage{float}
\usepackage{soul, color} 
\soulregister\ref{7}  
\soulregister\cite{7} 
\renewcommand{\st}[1]{}

\newcommand{\rv}{\mathbf{r}}

\newcommand{\appropto}{\mathrel{\vcenter{
  \offinterlineskip\halign{\hfil$##$\cr
    \propto\cr\noalign{\kern.2pt}\sim\cr\noalign{\kern-2.5pt}}}}}

\renewcommand{\Im}{\operatorname{Im}}
\newcommand{\tr}{\operatorname{Tr}}

\newcommand{\Pv}{\mathbf{P}}

\newcommand{\Ev}{\mathbf{E}}
\newcommand{\Dv}{\mathbf{D}}

\newcommand{\curl}{\nabla\times}
\newcommand{\inflim}{_{-\infty}^\infty}

\begin{document}
\rmfamily

\title{Casimir light in dispersive nanophotonics}
\author{Jamison Sloan$^{1}$, Nicholas Rivera$^{2}$, John D. Joannopoulos$^{2}$, and Marin Solja\v{c}i\'{c}$^{2}$}

\affiliation{$^{1}$ Department of Electrical Engineering and Computer Science, Massachusetts Institute of Technology, Cambridge, MA 02139, United States \\
$^{2}$ Department of Physics, Massachusetts Institute of Technology, Cambridge, MA 02139, United States}

\noindent	

\begin{abstract}

Time-varying optical media, whose dielectric properties are actively modulated in time, introduce a host of novel effects in the classical propagation of light, and are of intense current interest. In the quantum domain, time-dependent media can be used to convert vacuum fluctuations (virtual photons) into pairs of real photons. We refer to these processes broadly as ``dynamical vacuum effects'' (DVEs). Despite interest for their potential applications as sources of quantum light, DVEs are generally very weak, providing many opportunities for enhancement through modern techniques in nanophotonics, such as using media which support excitations such as plasmon and phonon polaritons. Here, we present a theory of DVEs in arbitrary nanostructured, dispersive, and dissipative systems. A key element of our framework is the simultaneous incorporation of time-modulation and ``dispersion'' through time-translation-breaking linear response theory. We propose a highly efficient scheme for generating entangled surface polaritons based on time-modulation of the optical phonon frequency of a polar insulator. We show that the high density of states, especially in hyperbolic polaritonic media, may enable high-efficiency generation of entangled phonon-polariton pairs. More broadly, our theoretical framework enables the study of quantum light-matter interactions in time-varying media, such as spontaneous emission, and energy level shifts.
\end{abstract}

\maketitle

The nonvanishing zero-point energy of quantum electrodynamics leads to a variety of observable consequences such as atomic energy shifts \cite{bethe1947electromagnetic}, spontaneous emission \cite{purcell1946purcell, gerard1999strong}, forces \cite{lamoreaux1997demonstration}, and non-contact friction \cite{kardar1999friction, pendry1997shearing}. Perhaps the most famously cited consequence of vacuum fluctuations is the Casimir effect \cite{mohideen1998precision, klimchitskaya2009casimir, bordag2001new, plunien1986casimir}, which predicts that two uncharged conducting plates, when placed close together, experience mutual attraction (or repulsion, in some cases \cite{munday2009measured, kenneth2002repulsive, zhao2009repulsive}) due to the fluctuating electromagnetic fields between the plates. The character of any fluctuation-based phenomenon is determined by the electromagnetic modes which exist around the structure of interest. As a result, the last two decades have provided promising insights about how nanostructured composites of existing and emerging optical materials can be used to modify observable effects of zero-point fluctuations. 

In time-varying systems, electromagnetic vacuum fluctuations can lead to the production of real photons. Famously, the ``dynamical Casimir effect'' predicts how a cavity with rapidly oscillating boundaries produces entangled photon pairs \cite{moore1970quantum}. Other related phenomena include photon emission from rotating bodies \cite{maghrebi2012spontaneous}, spontaneous parametric down-conversion in nonlinear materials \cite{boyd2019nonlinear}, the Unruh effect for relativistically accelerating bodies  \cite{yablonovitch1989accelerating, crispino2008unruh, fulling1976radiation, unruh1984happens}, Hawking radiation from black holes \cite{hawking1975particle, unruh1976notes}, and even particle production in the early universe \cite{shtanov1995universe}. The close connections among these phenomena are discussed in \cite{nation2012colloquium}. These ``dynamical vacuum effects'' (DVEs) have been studied in depth since the 1960s for their relation to fundamental questions about the quantum vacuum, and for their potential applications as quantum light sources \cite{glauber1991quantum, walls2007quantum, scully1999quantum}. Specifically, these processes are known to produce squeezed light (which is entangled if more than one mode is involved) \cite{loudon1987squeezed, breitenbach1997measurement} which enjoys applications in quantum information \cite{ralph1998teleportation}, spectroscopy \cite{polzik1992spectroscopy}, and enhancing phase sensitivity at LIGO \cite{aasi2013enhanced}. Despite high interest, these DVEs are very weak, with the first direct observation of the dynamical Casimir effect occurring as recently as 2011 \cite{wilson2011observation}. The strength of these effects can, in theory, be enhanced by nanostructured optical composites, and polaritonic materials with strong resonances, as has been seen with other fluctuation-based phenomena \cite{purcell1946purcell, rodriguez2011casimir, volokitin2007near}. However, considering DVEs in such materials is complicated by the subtleties associated with describing the optical properties of materials which are simultaneously dispersive and time-dependent. Beyond this fundamental issue, there is not yet a general framework which describes these emission effects in arbitrary nanostructured materials \cite{dodonov2010current}. Such a framework is of paramount importance if modern material and nanofabrication platforms are to be used to optimize these effects to make them practical for potential applications in quantum information, spectroscopy, imaging, and sensing.

In this Letter, we present a theoretical framework, based on macroscopic quantum electrodynamics (MQED), for describing DVEs in arbitrary nanostructured, dispersive, and dissipative time-dependent systems. We apply our theory to describe two-photon emission processes from time-varying media. As an example, we show that phonon-polariton pairs can be generated on thin films of polar insulators (e.g., silicon carbide and hexagonal boron nitride), whose transverse optical (TO) phonon frequency is rapidly modulated in time. We find that the high density of states of surface phonon-polariton modes, in conjunction with dispersive resonances, leads to phonon-polariton pair generation efficiencies which are orders of magnitude higher than traditional parametric down conversion. Our results are particularly relevant in the context of recent experiments, which have observed parametric amplification of optical phonons in the presence of a strong driving field, which effectively causes the TO phonon frequency to vary in time \cite{cartella2018parametric}.

\begin{figure}[t]
    \centering
    \includegraphics[width=\linewidth]{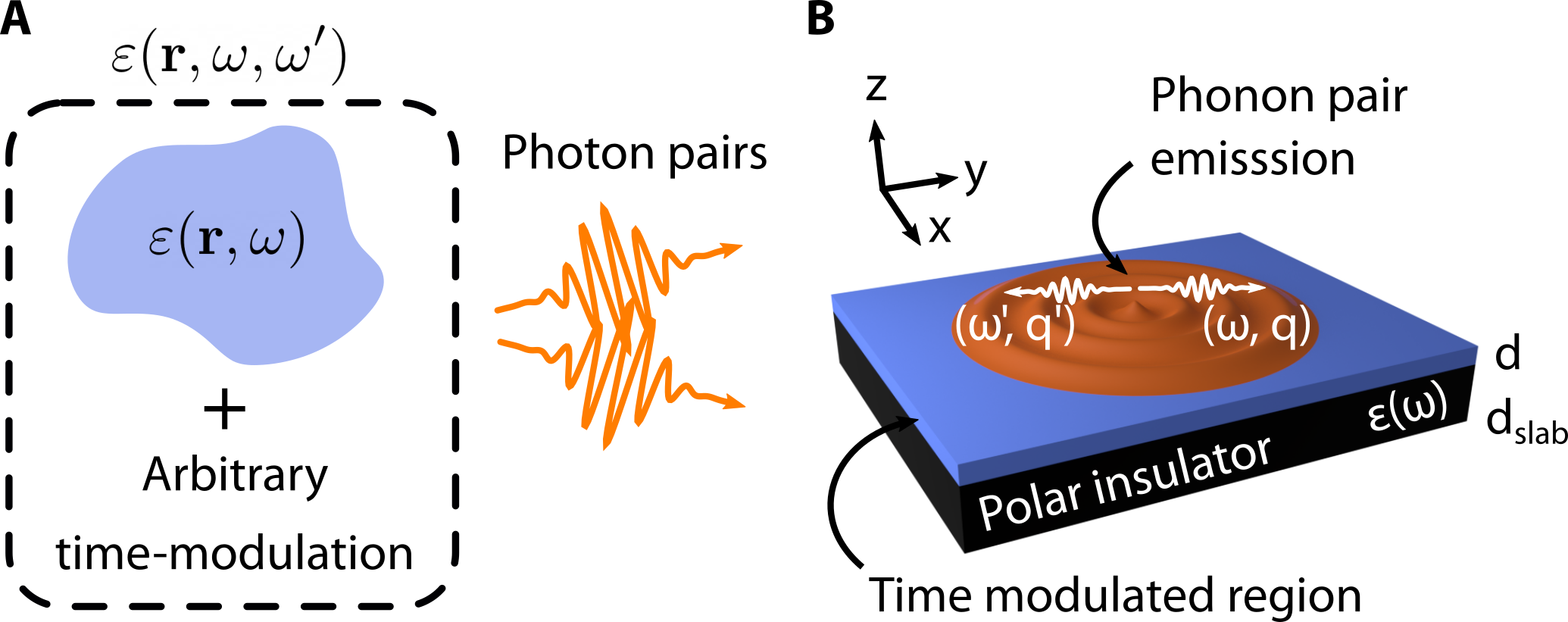}
    \caption{\textbf{Photon pair emission from arbitrary time-dependent dielectric media.} (a) A dispersive dielectric $\varepsilon(\rv,\omega)$ subject to an arbitrary time modulation can be described as having the more general dielectric function $\varepsilon(\rv, \omega, \omega')$ which encodes both time dependence and dispersion. (b) A schematic of a thin film of polar insulator which has a small top layer which undergoes a time modulation. As a result, surface phonon-polariton pairs are produced with frequencies $\omega, \omega'$, and wavevectors $q, q'$.}
    \label{fig:schematic}
\end{figure}

There are inherent subtleties associated with describing the optical response of time-modulated dielectrics which are already dispersive. In systems where frequencies of time-modulation are far from any transition frequencies in the system, one can consider an ``adiabatic'' description of the time-dependent material. In this case, the permittivity can be taken as $\varepsilon(\omega; t)$, or simply $\varepsilon(t)$, as is done in many theoretical and experimental studies \cite{law1994effective, lustig2018topological, zurita2009reflection,chu1972wave,harfoush1991scattering,fante1971transmission,holberg1966parametric}. In cases where the adiabatic approximation breaks down (e.g. in disperisve systems with similar modulation and transition frequencies), we must revert to the most general dielectric function allowed by linear response theory. In the absence of time-translation invariance, the polarization $\mathbf{P}(t)$ is connected to the applied field $\Ev(t)$ through a susceptibility $\chi(t,t')$. Consequently, the frequency response must be characterized by a two-frequency susceptibility $\chi(\omega,\omega') \equiv \int\inflim dt\,dt'\, \chi(t,t') e^{i\omega t}e^{-i\omega't'}.$ The two Fourier transforms are defined with opposing sign conventions so that for a time-independent material, $\chi(\omega, \omega') = 2\pi\delta(\omega - \omega')\chi(\omega)$. The corresponding permittivity is defined by $\varepsilon(\omega,\omega') = 2\pi\delta(\omega-\omega') + \chi(\omega,\omega')$, as illustrated in Fig. \ref{fig:schematic}a. In this case, the displacement field $\Dv$ is connected to the electric field $\Ev$ as
\begin{equation}
    \mathbf{D}(\omega) = \int\inflim \frac{d\omega'}{2\pi}\varepsilon(\omega,\omega')\Ev(\omega').
    \label{eq:nonlocal_constitutive}
\end{equation}
We now use this apparatus to parameterize the types of time-modulated materials we consider in our theory of DVEs. Consider a photonic structure (of arbitrary geometry and material composition), with a local dispersive dielectric function $\varepsilon_{\text{bg}}(\rv, \omega)$. Then we impart some spatiotemporal change to the susceptibility $\Delta\chi(\rv, \omega, \omega')$, so that the total permittivity is
\begin{equation}
    \varepsilon(\rv,\omega,\omega') = \varepsilon_{\text{bg}}(\rv, \omega)[2\pi\delta(\omega-\omega')] + \Delta\chi(\rv, \omega, \omega').
    \label{eq:epsilon_perturbation}
\end{equation}

Our theory of DVEs in systems described by the general form of Eq. \ref{eq:epsilon_perturbation} is based on a Hamiltonian description of electromagnetic field subject to interactions in general time-varying media. We use macroscopic quantum electrodynamics (MQED) \cite{scheel2009macroscopic, rivera2020light} to quantize the electromagnetic field in the background structure $\varepsilon_{\text{bg}}(\rv, \omega)$. In this framework, the Hamiltonian of the bare electromagnetic field is 
\begin{equation}
    H_{\text{EM}} = \int_0^\infty d\omega \int d^3r\, \hbar\omega\, \mathbf{f}^\dagger(\rv,\omega)\cdot\mathbf{f}(\rv,\omega),
\end{equation}
where $\mathbf{f}^{(\dagger)}(\rv,\omega)$ is the annihilation (creation) operator for a quantum harmonic oscillator at position $\rv$ and frequency $\omega$. In such a medium, the electric field operator in the interaction picture is given as 
\begin{equation}
\begin{split}
    \Ev(\rv, t) = i \sqrt{\frac{\hbar}{\pi\varepsilon_0}} \int_0^\infty &d\omega \frac{\omega^2}{c^2} \int d^3r' \sqrt{\Im \varepsilon_{\text{bg}}(\rv', \omega)} \\
    &\times\left(\mathbf{G}(\rv,\rv',\omega)\mathbf{f}(\rv',\omega)e^{-i\omega t} - \text{h.c.}\right).
\end{split}
\label{eq:efield_operator}
\end{equation}
Here, $\mathbf{G}(\rv,\rv',\omega)$ is the electromagnetic Green's function of the background which satisfies $\left(\curl\curl - \varepsilon_{\text{bg}}(\rv,\omega)\frac{\omega^2}{c^2}\right)\mathbf{G}(\rv,\rv',\omega) = \delta(\rv-\rv')I$, where $I$ is the $3\times 3$ identity matrix. We assume that the permittivity change described by Eq. \ref{eq:epsilon_perturbation} creates a change to the polarization density $\Pv(\rv, t)$, interacting with the electric field via $V(t) = -\int d^3r\,\mathbf{P}(\rv,t)\cdot\mathbf{E}(\rv,t)$ \cite{boyd2019nonlinear}. Then by relating the polarization to the electric field through linear response, we find the interaction Hamiltonian 
\begin{equation}
    V(t) = -\varepsilon_0 \int d^3r\,dt'\,\Delta\chi_{ij}(\rv, t, t')E_j(\rv,t') E_i(\rv,t),
    \label{eq:interaction_H}
\end{equation}
where we have used repeated index notation. If we work in the regime where $\Delta\chi$ is small, then the electric field operator is well-approximated by that of the unperturbed field, given in Eq. \ref{eq:efield_operator}. To compute rates of two-photon emission, we consider scattering matrix elements that connect the electromagnetic vacuum state to final states which contain two photons. Taking the S-matrix elements to first order in perturbation theory (see S.I.), the probability of two-photon emission is given by
\begin{equation}
    \begin{split}
    P = \frac{1}{2\pi^2 c^4}&\int_0^\infty d\omega\,d\omega'\,(\omega\omega')^2 \int d^3r\,d^3r'  \\
    &\times\tr\left[\Delta\chi(\rv, \omega,-\omega')\Im G(\rv,\rv',\omega')\right. \\
    &\hspace{1cm}\Delta\chi^\dagger(\rv', \omega,-\omega')\Im G(\rv',\rv,\omega)\left.\right],
    \end{split}
    \label{eq:master_probability}
\end{equation}
where $\Delta\chi^\dagger$ is the matrix conjugate transpose of the tensor $\Delta\chi$. The two instances of $\Im G$ indicates that there are two quanta emitted --- one at frequency $\omega$, and the other at $\omega'$. The Green's function encodes everything about the structure, dispersion, and dissipation of the background structure, and its imaginary part is closely related to the local density of states, suggesting that emission probabilities can be increased when more modes are available, much as in the well-known case of single-photon Purcell enhancement. Similar results have also been seen with two-photon emission from atoms \cite{rivera2017making}. Meanwhile, the tensor $\Delta\chi$ encodes everything about the imposed time dependence of the material. 
This separation makes the computation of emission rates in photonic nanostructures highly modular, and may provide future opportunities for numerical implementations in cases where analytical results are not feasible.

\begin{figure}[ht]
    \centering
    \includegraphics[width=\linewidth]{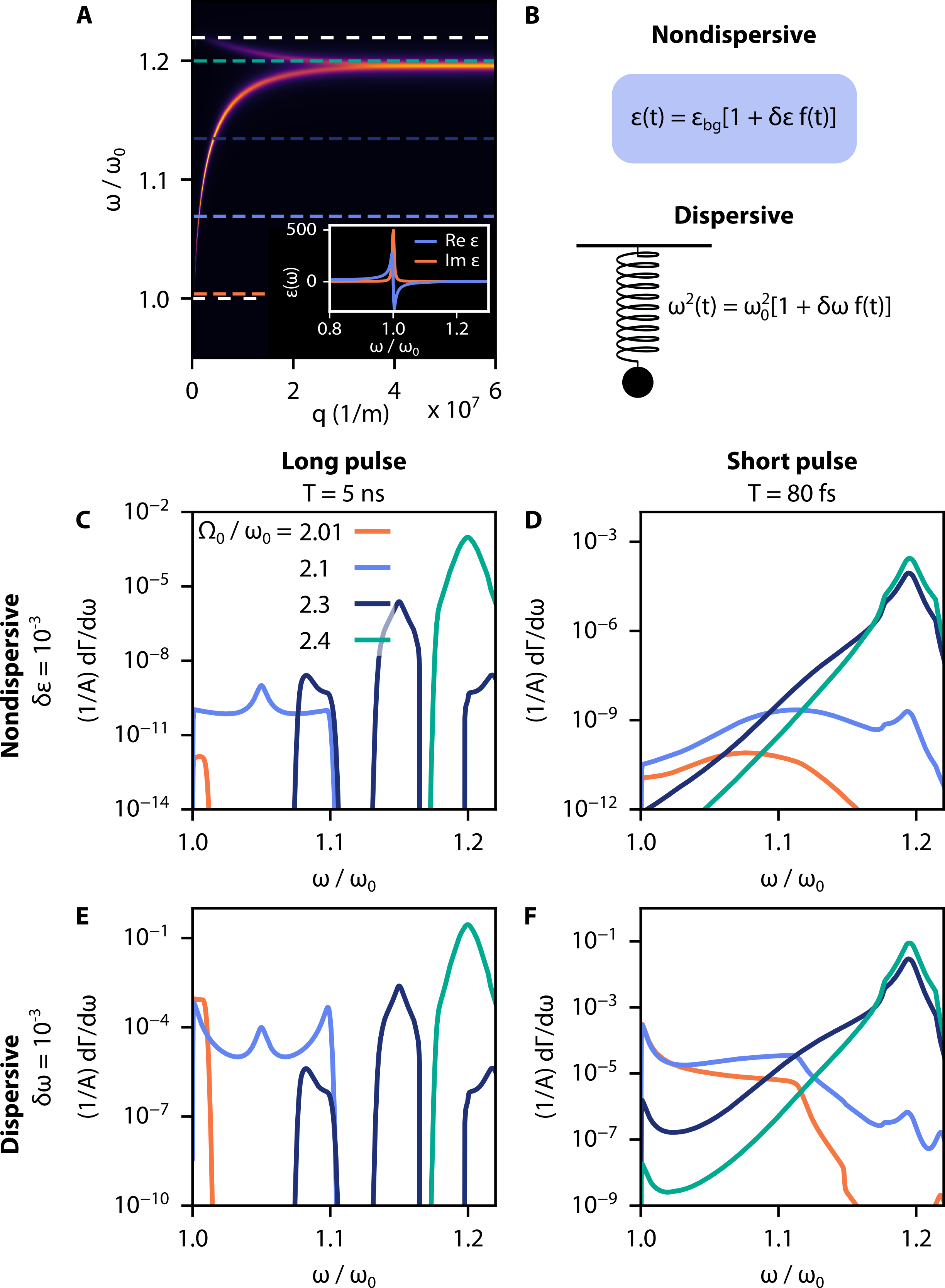}
    \caption{\textbf{Dynamical Casimir effect for silicon carbide phonon-polaritons.} (a) Dispersion relation of phonon-polaritons on $d_{\text{slab}} = 100$ nm thick slab of SiC. Dotted lines mark the edges of the RS band. Inset shows the Lorentz oscillator permittivity around $\omega_0 = 1.49\times 10^{14}$ rad/s. (b) Schematic representation of nondispersive time modulations of the permittivity, versus dispersive modulations of the transverse optical phonon frequency $\omega_0$. (c-f) Differential rate per unit area $(1/A) d\Gamma/d\omega$ for phonon-polariton pairs production for various values of $\Omega_0/\omega_0 = \{2.01, 2.1, 2.3, 2.4\}$, as well as short ($T= 80$ fs) and long ($T = 5$ ns) pulses. The modulated region is assumed to be $d=10$ nm thick. Panels (c, d) show a nondispersive modulation with $\delta\varepsilon = 10^{-3}$. Panels (e, f) show a dispersive modlation with $\delta\omega = 10^{-3}$.}
    \label{fig:SiC_DCE}
\end{figure}

We now show how our theoretical framework accounts for dispersion and loss in time-modulated thin films which generate pairs of entangled surface polaritons. Surface polaritons have enjoyed a myriad of applications due to their ability to maintain high confinement, and relatively low loss \cite{chen2012optical, dai2014tunable, dai2015graphene, basov2016polaritons}. Specifically, we examine surface phonon-polaritons (SPhPs) on thin films of the polar insulators silicon carbide (SiC) and hexagonal boron nitride (hBN). Controllable sources of entangled surface polaritons are notably lacking, and could prove important for applications in quantum information and imaging. In the infrared, the dielectric response of polar insulators is well-described by the resonance of transverse optical (TO) phonon modes. The permittivity in this frequency range is given by the Lorentz oscillator
$
    \varepsilon_{\text{bg}}(\omega) = \varepsilon_{\infty} + \omega_p^2/(\omega_0^2 - \omega^2 - i\omega\Gamma),
$
where $\varepsilon_\infty$ is the permittivity at high frequencies, $\omega_0$ is the TO phonon frequency, $\omega_p$ is the plasma frequency, and $\Gamma$ is the damping rate. Phonon-polaritons are supported above the resonance at $\omega_0$, where $\text{Re}\,\varepsilon_{\text{bg}}(\omega) < -1$, which is referred to as the Reststrahlen band, or ``RS band'' (Fig. \ref{fig:SiC_DCE}a). 

To highlight the interplay between dispersion and time dependence in two-polariton spontaneous emission, we compare two different modulations of the polar insulator structures (Fig. \ref{fig:SiC_DCE}b). The first is a nondispersive modulation, where a layer of thickness $d$ has its index perturbed by a constant amount as $\varepsilon(t) = \varepsilon_{\text{bg}}(1 + \delta\varepsilon\,f(t))$. In this case, we have $\Delta\chi(\omega,\omega') = \delta\varepsilon\,f(\omega-\omega')$, where $f(\omega)$ is the Fourier transform of the modulation profile. If the change in index is caused by a nonlinear layer with $\chi^{(2)} = 100$ pm/V, then an electric field strength of $10^7$ V/m gives $\delta\varepsilon = 10^{-3}$. The second is a dispersive modulation, where over a thickness $d$, the transverse optical phonon frequency $\omega_0$ is modulated to deviate from its usual value as a function of time as $\omega^2(t) = \omega_0^2(1 + \delta\omega\,f(t))$. In this case, $\Delta\chi(\omega, \omega') = \delta\omega\,\omega_0^2 \omega_p^2 f(\omega-\omega')/(Q(\omega)Q(\omega'))$ to first order in $\delta\omega$, where $Q(\omega) \equiv \omega_0^2 - \omega^2 - i\omega\Gamma$ (see S.I.). From the experimental models presented in \cite{cartella2018parametric} for SiC, we estimate that an applied field strength of 1 GV/m gives rise to a frequency shift of the order $\delta\omega = 10^{-3}$. We will compare the two modulation types with the same fractional change in parameter $\delta\varepsilon = \delta\omega = 10^{-3}$ to highlight that around $\omega_0$, a fractional change $\delta\omega$ causes much stronger effects than $\delta\varepsilon$. Later, we comment on efficiencies given the same applied field strength.

\begin{figure*}[ht]
    \centering
    \includegraphics[width=\textwidth]{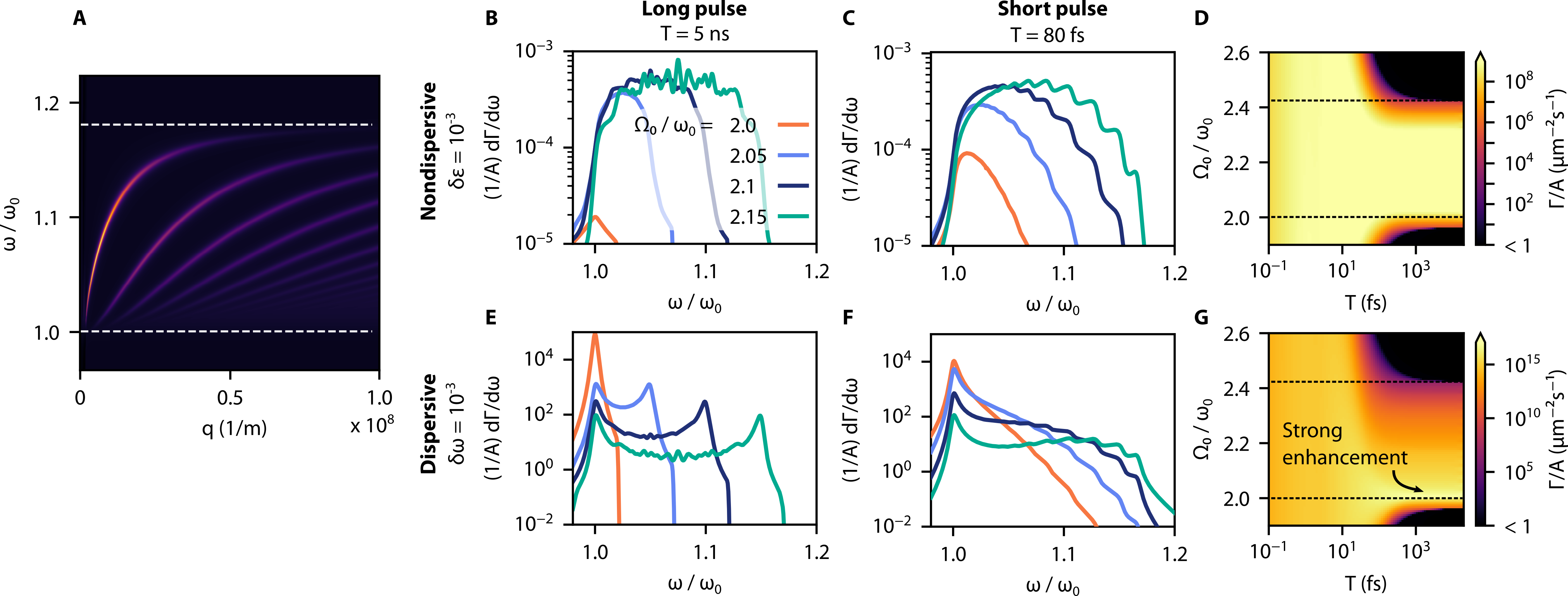}
    \caption{\textbf{Achieving strong DVE s through dispersive modulations.} (a) Dispersion of surface phonon-polaritons on a thin layer of hBN ($d_{\text{slab}} = 100$ nm) in the upper RS band ($\omega_0 = 2.56\times 10^{14}$ rad/s). (b, c) Differential rate per unit area $(1/A) d\Gamma/d\omega$ for phonon-polariton pairs production for various values of $\Omega_0/\omega_0 = \{2, 2.05, 2.1, 2.15\}$, as well as short ($T= 80$ fs) and long ($T = 5$ ns) pulses. (d) Total emission rate per area of phonon-polariton pairs as a function of pulse duration $T$ and frequency $\Omega_0$. (e-g) Same as (b-d), except that the modulation is dispersive. Panel (g) shows the strong enhancement which occurs for monochromatic modulations when $\Omega_0/\omega_0 = 2$, corresponding to enhancement of DVEs by dispersive parametric amplification.}
    \label{fig:DCE_hBN}
\end{figure*}

We modulate the surface layer with perturbations of the form $f(t) = \cos(\Omega_0 t)e^{-t^2/2T^2}$. 
This enables us to consider modulations across many timescales, from ultrashort pulses, to nearly monochromatic (CW) modulations. Applying our formalism to the geometry depicted in Fig. \ref{fig:schematic}b, we find that the probability of two-polariton emission per unit frequency $\omega$ and $\omega'$ is given as
\begin{equation}
\begin{split}
    \frac{1}{A}\frac{dP}{d\omega d\omega'} = \frac{|\Delta\chi(\omega, -\omega')|^2}{16\pi^3}\int_0^\infty dq\, q \left(1 - e^{-2qd} \right)^2 \\
    \times\Im r_p(\omega, q) \Im r_p(\omega', q).
\end{split}
    \label{eq:phonon_DCE_general_differential}
\end{equation}
Here, $r_p(\omega, q)$ is the p-polarized reflectivity associated with the interface, and $A$ is the sample area. This equation encodes the frequency correlations between the two emitted quanta $\omega$ and $\omega'$. Once $\Delta\chi$ is chosen, Eq. \ref{eq:phonon_DCE_general_differential} can be integrated over $\omega'$ and normalized by the pulse duration $T$ to obtain an area-normalized rate per frequency $(1/A) d\Gamma/d\omega$. This quantity represents the emission rate which is detected classically at frequency $\omega$, and thus no longer discriminates between the two photons of the emitted pair. 

Using this method, we obtain results for SiC which is modulated both dispersively and nondispersively. In Fig. \ref{fig:SiC_DCE}a, we see the phonon-polariton dispersion for a 100 nm layer of SiC (dielectric parameters taken from \cite{le1997experimental}) Figs. \ref{fig:SiC_DCE}c-f show the corresponding rate distribution $(1/A) d\Gamma/d\omega$ for each of the marked modulation frequencies. We first note the difference between nearly monochromatic modulations and short pulses. For a long pulse (Fig. \ref{fig:SiC_DCE}c), the two emitted polaritons are subject to the energy conservation constraint $\omega + \omega' \approx \Omega_0$. In this regime, the behavior of the rate spectrum $d\Gamma/d\omega$ is determined by where $\Omega_0/2$ lies in the RS band (see dashed lines on Fig. \ref{fig:SiC_DCE}a). We see that for various $\Omega_0$, the spectra are symmetrically peaked around $\Omega_0/2$, with widths set by the loss. The strongest response is seen around $\Omega_0/\omega_0 = 2.4$ where the density of states of SPhPs is highest. At the slightly lower excitation frequency $\Omega_0/\omega_0 = 2.3$, the central peak at $\Omega_0/2$ is flanked by two symmetrical side peaks. These secondary peaks occur since $\Omega_0/2$ lies in between two bands of the dispersion, and thus one possibility for satisfying the approximate energy conservation relation is that one polariton is emitted into each band at the same wavevector $q$. Also notably, we see that the modulation associated with $\Omega_0/\omega_0 = 2.01$ produces very little response, owing to the low density of states at the bottom of the RS band. For a short pulse (Fig. \ref{fig:SiC_DCE}), the general trend in magnitudes between the excitation frequencies is the same. However, maximum rate that can be achieved is 10-100 times smaller, as the pulse is not long enough to establish a well-defined frequency. Additionally, since a short pulse eliminates the strict energy conservation condition, polaritons can be emitted at many frequency pairs. As a result, the shape of the spectrum for most excitation frequencies is peaked near the top of the RS band where the density of states is highest. 

For dispersive modulations, many aspects of SPhP pair production remain the same. However, several key changes emerge as a result of the difference in the factor $|\Delta\chi|^2 \propto 1/|Q(\omega)Q(\omega')|^2$, which becomes large when $\omega, \omega' \approx \omega_0$. This condition corresponds to parametric resonance of phonons which dictate the dielectric response. While the behavior of the monochromatic modulation (Fig. \ref{fig:SiC_DCE}e) for higher frequencies $\Omega_0$ remains qualitatively the same, the magnitudes of the peaks for $\Omega_0/\omega_0 = 2.1, 2.01$ increase substantially. Interestingly, for $\Omega_0/\omega_0 = 2.1$, this resonance amplifies the tails of the frequency distribution, so that nondegenerate pair production is actually slightly preferred. For short pulses (Fig. \ref{fig:SiC_DCE}f), the density of states behavior remains largely unchanged. However, the tails of the distribution at the bottom of the RS band near $\omega_0$ are raised, in contrast to the nondispersive behavior (Fig. \ref{fig:SiC_DCE}d). There are two main factors which may cause strong enhancement of the phonon emission spectrum: high density of states, and parametric resonance around $\omega_0$. For SiC, these large dispersive enhancements occur around $\omega_0$, which is actually at a point of very low density of states in the dispersion. We can then reason that the strongest emission should come from systems where the dispersive resonance overlaps more strongly with the high density of states.

To this end, we elucidate how SPhPs on hBN, due to their multi-banded nature, can enjoy much stronger enhancement through dispersive modulations. Unlike SiC, hBN is an anisotropic polar insulator, with different transverse optical phonon frequencies in the in-plane or out-of-plane directions. As a result, hBN has two RS bands, and the dispersion relation is hyperbolic, being multi-branched in each RS band \cite{basov2016polaritons}. The dispersion relation in the RS band of thin hBN is seen in Fig. \ref{fig:DCE_hBN}a (dielectric parameters taken from \cite{woessner2015highly, cai2007infrared}). In contrast to SiC, the density of states of SPhPs is spread broadly across the upper RS band. Figs. \ref{fig:DCE_hBN}b,c show the emitted pair spectrum for a variety of driving frequencies, similarly to SiC. The fringes seen in the emission spectra are a direct consequence of interference between many possibilities for how two phonon-polaritons can distribute themselves into many branches of the dispersion. Fig. \ref{fig:DCE_hBN}d shows the total rate of emission integrated over the upper RS band for a range of modulation frequencies $\Omega_0$ and pulse durations $T$. Due to the relatively even density of states, we see that the emission rate in the nondispersive case is relatively uniform ($\Gamma/A \approx 10^9 \mu$m$^{-2}$s$^{-1}$) across a wide range of parameters.  

For a dispersive modulation, the emission strengths are reordered entirely, and the strongest emission occurs for degenerate production around $\omega_0$ when the system is modulated at $2\omega_0$. These differences manifest not only in the shape of the spectrum, but in the overall strength of each process. Specifically, we see that around the point of strongest enhancement (Fig. \ref{fig:DCE_hBN}g), the emission rate is orders of magnitude higher than for long pulses outside of the resonance around $\omega_0$. Even though achieving $\delta\varepsilon = 10^{-3}$ through a nonlinear substrate requires a lower applied field than a TO phonon frequency shift of equivalent proportion, the sensitive nature of the dispersive modulations provides opportunities for improved efficiency. We estimate that with an applied field strength of 1 GV/m, the nondispersive modulation achieved through a thin nonlinear ($\chi^{(2)} = 100$ pm/V) layer has a quantum efficiency of the order $\eta \approx 10^{-9}$, while at the same field strength, the dispersive modulation has $\eta \approx 10^{-5}$. Given that evidence of parametric amplification of optical phonons in SiC has already been demonstrated \cite{cartella2018parametric}, we believe that efficient generation of SPhP pairs on SiC and hBN by optical excitation should be feasible. We have also applied our formalism to the generation of graphene plasmons on a nonlinear substrate, and found this process could have an efficiency $\eta \approx 10^{-4}$ (see S.I.). Such efficiencies could exceed the highest seen for pair generation to date \cite{bock2016highly}. Potential applications will need to use or out-couple SPhP pairs before they are attenuated.

We have provided a comprehensive Hamiltonian theory which governs photon interactions in dispersive time-dependent dielectrics. Our work shows that the role of dispersion is critical in describing and enhancing these phenomena, as we showed for time-modulated polar insulators. Our framework is amenable to design and optimization of complex structures for experiments and potential devices. Our theory may also provide important insights about how enhanced nonlinearities in epsilon-near-zero materials \cite{caspani2016enhanced, alam2016large} may present opportunities for enhancing DVEs by realizing large relative changes in the permittivity. Beyond this, the Hamiltonian MQED formalism we have presented can enable further studies of light-matter interactions in arbitrary time-dependent materials. For example, one could model how spontaneous emission and energy-level shifts of quantum emitters are modified in the presence of time-modulation. Finally, this kind of formalism could provide opportunities for studying the role that parametric amplification of quasiparticles can play in exotic effects in solid-state systems such as light-induced superconductivity \cite{mitrano2016possible, babadi2017theory}. Broadly, we anticipate that our framework will be of interest for describing classical and quantum phenomena in many timely experimental platforms featuring ultrafast optical modulation of materials.

\begin{acknowledgements}
    The authors thank Yannick Salamin and Prof. Ido Kaminer for helpful discussions. This material is based upon work supported by the Defense Advanced Research Projects Agency (DARPA) under Agreement No. HR00112090081. This work was supported in part by the U.S. Army Research Office through the Institute for Soldier Nanotechnologies under award number W911NF-18-2-0048. J.S was supported in part by NDSEG fellowship No. F-1730184536. N.R. was supported by Department of Energy Fellowship DE-FG02-97ER25308. 
\end{acknowledgements}

\bibliography{vacuum}
\newpage
\onecolumngrid
\appendix 

\newpage

\end{document}